\documentclass[12pt,preprint]{aastex}

\shorttitle{Stellar Rotation in NGC\,6811 with {\it Kepler}}
\shortauthors{Meibom et al.}

\begin{document}

\title{THE KEPLER CLUSTER STUDY: \\
STELLAR ROTATION IN NGC\,6811}

\author{S{\o}ren Meibom\altaffilmark{1},
Sydney A. Barnes\altaffilmark{2},
David W. Latham\altaffilmark{1},
Natalie Batalha\altaffilmark{3},
William J. Borucki\altaffilmark{3},
David G. Koch\altaffilmark{3},
Gibor Basri\altaffilmark{4},
Lucianne M. Walkowicz\altaffilmark{4},
Kenneth A. Janes\altaffilmark{5},
Jon Jenkins\altaffilmark{6,3},
Jeffrey Van Cleve\altaffilmark{6,3},
Michael R. Haas\altaffilmark{3},
Stephen T. Bryson\altaffilmark{3},
Andrea K. Dupree\altaffilmark{1},
Gabor Furesz\altaffilmark{1},
Andrew H. Szentgyorgyi\altaffilmark{1},
Lars A. Buchhave\altaffilmark{7,1},
Bruce D. Clarke\altaffilmark{6},
Joseph D. Twicken\altaffilmark{6},
Elisa V. Quintana\altaffilmark{6}}

\altaffiltext{1}{Harvard-Smithsonian Center for Astrophysics, Cambridge,
MA, 02138, USA}
\altaffiltext{2}{Lowell Observatory, Flagstaff, AZ, 86001, USA}
\altaffiltext{3}{NASA Ames Research Center, Moffett Field, CA 94035, USA}
\altaffiltext{4}{Astronomy Department, University of California, Berkeley,
CA 94720, USA}
\altaffiltext{5}{Department of Astronomy, Boston University, Boston,
MA 02215, USA}
\altaffiltext{6}{SETI Institute, Mountain View, CA 94043, USA}
\altaffiltext{7}{Niels Bohr Institute, Copenhagen University, Denmark}

\begin{abstract}
We present rotation periods for 71 single dwarf members of the open
cluster NGC\,6811 determined using photometry from NASA's {\it Kepler}
Mission. The results are the first from {\it The Kepler Cluster
Study} which combine {\it Kepler}'s photometry with ground-based
spectroscopy for cluster membership and binarity. The rotation periods
delineate a tight sequence in the NGC\,6811 color-period diagram from
$\sim$1\,day at mid-F to $\sim$11\,days at early-K spectral type. This
result extends to $\sim$1\,Gyr similar prior results in the $\sim$600\,Myr
Hyades and Praesepe clusters, suggesting that rotation periods for cool
dwarf stars delineate a well-defined surface in the 3-dimensional space
of color (mass), rotation, and age. It implies that reliable ages can
be derived for field dwarf stars with measured colors and rotation
periods, and it promises to enable further understanding of various
aspects of stellar rotation and activity for cool stars.
\end{abstract}



\section{INTRODUCTION}
\label{intro}

Cool stars lose angular momentum and spin down with time. Observations
indicate that the surface rotation period $P$, of a cool main sequence
star\footnote{With the exception of stars in systems where tidal effects
have influenced the rotation rate.} is mainly dependent on its age $t$,
and mass $M$; $P = P(t, M)$.  
These dual dependencies suggest that a surface exists in the $P$-$t$-$M$
space which can be defined from measurements of the colors (masses) and
periods for stars with known ages. Observations to define the $P$-$t$-$M$
surface will simultaneously provide the dependence of $P$ on $M$ at a
given $t$ (cross section across $t$-axis) and the dependence of $P$ on $t$
for a given $M$ (cross section across $M$-axis). These mass- and
age-dependencies of stellar spin-down will enable a more detailed
understanding of the physical processes behind the loss of angular momentum
in cool stars \citep[e.g.][]{kawaler88,pkd90,ssm+93,cl94,bfa97,dpt+10,bk10}.
The existence of a well-defined surface in $P$-$t$-$M$ space will also imply
that the measurement of two of these variables yields the third. Of the
three variables, $P$ and $M$ (or a suitable proxy such as color) are the
easiest to measure, providing access to the stellar age ($t$), invaluable
for the chronological arrangement of stars and their companions (including
planets) and thus for our understanding of various related astrophysical
phenomena. The thickness of the $P$-$t$-$M$
surface will determine how well the theory of angular momentum evolution
can be constrained and how precisely ages can be determined.

Rotation period measurements in young open clusters are steadily
subsuming earlier $v \sin i$ data
\citep[e.g.][]{sh87,ssh+93,jfs96,sjf01,tsp+00,tpj+02}
and are helping to improve the definition of $P(t, M)$. For stars in
young ($\sim$100\,Myr) clusters the relationship between $P$ and $M$
is not unique, and the surface splits into two (fast and slow) branches
\citep[e.g.][]{barnes03a,hgp+09,hbk+10,mms09,mms+11,iab+09,jbm+10}.
With time, the fast branch dissipates as stars converge in their rotational
evolution onto the slow branch, resulting in a unique relationship between
$P$ and $M$ at $\sim$600\,Myr \citep{rtl+87,cdh+09,dch+11}. The Hyades
and Praesepe clusters constitute the oldest coeval stellar populations
for which rotation periods have been measured. Older stars rotate more
slowly and are less active, and the lack of periods in clusters older
than Hyades/Praesepe reflects the challenging task of measuring - from
the ground - the spot-induced photometric fluctuations required to
derive rotation periods. Consequently, the shape and position of the
$P$-$t$-$M$ surface for $t \ga 600$\,Myr is constrained only by
the rotation period, age, and mass of a single star - the Sun. 
(Ages of field stars with measured periods are too imprecisely known
to specify the surface.)

NASA's {\it Kepler} mission offers a special opportunity to overcome this
difficulty. {\it Kepler} provides nearly uninterrupted photometric measurements
of unprecedented duration ({\it years}), cadence ({\it minutes}),
and precision ({\it ppm}). As part of the {\it Kepler} mission, {\it The
Kepler Cluster Study} \citep{meibom10} is targeting three open
clusters older than the Hyades with the goal of deriving rotation periods
for their cool main-sequence members. If successful, this effort will
dramatically improve our empirical understanding of the spin-down rates
of low-mass stars of different masses by verifying the existence and
precisely defining the shape of the $P$-$t$-$M$ surface beyond the age of
the Hyades and possibly that of the Sun. In this paper we report the
first results from the Kepler Cluster Study for the $\sim$1\,Gyr
cluster NGC\,6811. The results confirm the existence of a unique
surface out to the age of NGC\,6811, and specify its shape, $P(M)$,
at that age. In future papers, we will extend $P(t, M)$ to $t = 2.5$\,Gyr
(NGC\,6819) and possibly $t = 9$\,Gyr (NGC\,6791).

\begin{figure}[!ht]
\epsscale{1.0}
\plotone{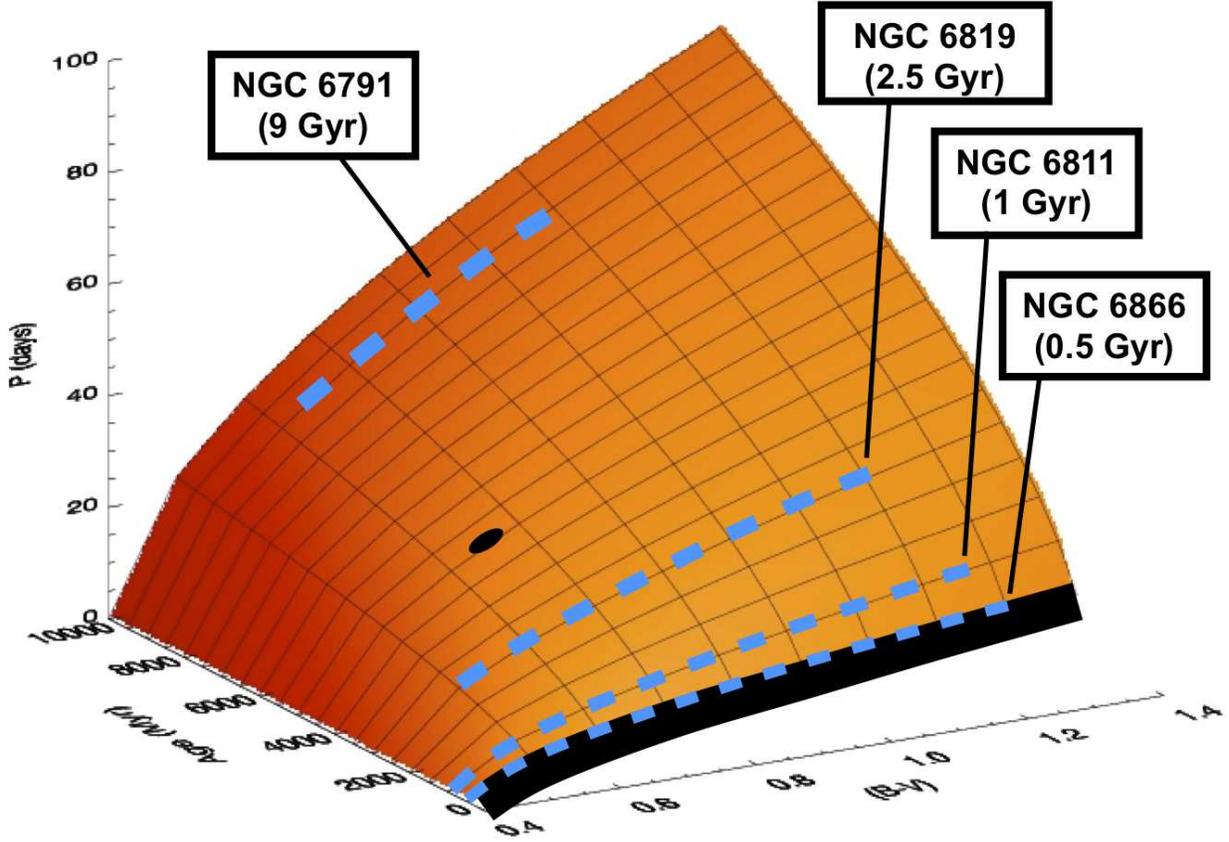}
\caption{
A surface (orange) in the 3-dimensional space of color (mass, x-axis),
age (Myr, y-axis), and stellar rotation period (Days, z-axis). The surface
is an extrapolation in age, using $P \propto \sqrt{t}$ \citep{skumanich72},
of the color-period relation observed among moderate-to-slow rotators in
the Hyades and younger clusters (black curve). The black dot marks the
color, age, and rotation period of the Sun. The dashed blue curves mark
the ages and color-ranges of the stars being observed by {\it Kepler}
in the four open clusters located within its field of view.
\label{3d}}
\end{figure}

\section{THE KEPLER CLUSTER STUDY}
\label{kocs}
Figure~\ref{3d} demonstrates the current severely uneven observational 
coverage of the $P$-$t$-$M$ parameter-space. The orange surface represents
an extrapolation in age, using the Skumanich $P \propto \sqrt{t}$ spin-down
law \citep{skumanich72}, of the color-period relation observed among
moderate-to-slow rotators in the Hyades and younger clusters (black curve). 

The Kepler Cluster Study is a program to identify members of the 
four open clusters within the {\it Kepler} field of view and to
obtain and analyze {\it Kepler} light curves for those members to measure
stellar rotation periods and search for transiting planets. The four
clusters are NGC\,6866 (0.5\,Gyr), NGC\,6811 (1\,Gyr), NGC\,6819
(2.5\,Gyr), and NGC\,6791 (9\,Gyr). As coeval, cospatial,
and chemically homogeneous collections of stars with a range of masses,
for which precise ages can be determined, open clusters are the best
opportunity we have for studying the dependencies of rotation on the
most fundamental stellar properties - age and mass. The potential
contributions to the study of stellar rotation by the Kepler 
Cluster Study are shown as blue dashed curves in Figure~\ref{3d}.
The curves represent the color ranges for the main-sequence members
currently being observed by {\it Kepler} in the four clusters.

Although certain information about NGC\,6811 is already available 
\citep{sanders71,lindoff72,bzs78,gbi99,mds+05,lzl+09}, 
it was not particularly well-studied prior to its inclusion in the
Kepler Cluster Study. Consequently, even basic properties of the
cluster are uncertain or unknown. Located near Cygnus and Lyra
($\alpha_{2000} = 19^{h}~37{m}$, $\delta_{2000} = +46\degr~23\arcmin$;
$l = 79\fdg2$, $b = 12\fdg0$), its color-magnitude diagram (CMD) is
highly contaminated with field stars (see Figure~\ref{cmd}), making
an extensive ground-based radial-velocity survey essential to identify
cluster members and to improve cluster parameters. The 1\,Gyr cluster
age quoted in this study is based on the recent photometric study by
\citet{mds+05} who found an age of 975\,Myr, and on an estimate of
the cluster age of $1.1 \pm 0.2$\,Gyr based on the color difference
between the main sequence turnoff and the red giant clump \citep{jh11}.
The latter technique is independent of the cluster reddening and only
moderately sensitive to the cluster metallicity. The current uncertainty
in the values for the reddening and metallicity of NGC\,6811 are the
limiting factors in a determination of its age from main-sequence and
turnoff fitting in the CMD.

\subsection{Ground-Based Spectroscopy}
\label{spec}
\begin{figure}[!ht]
\plotone{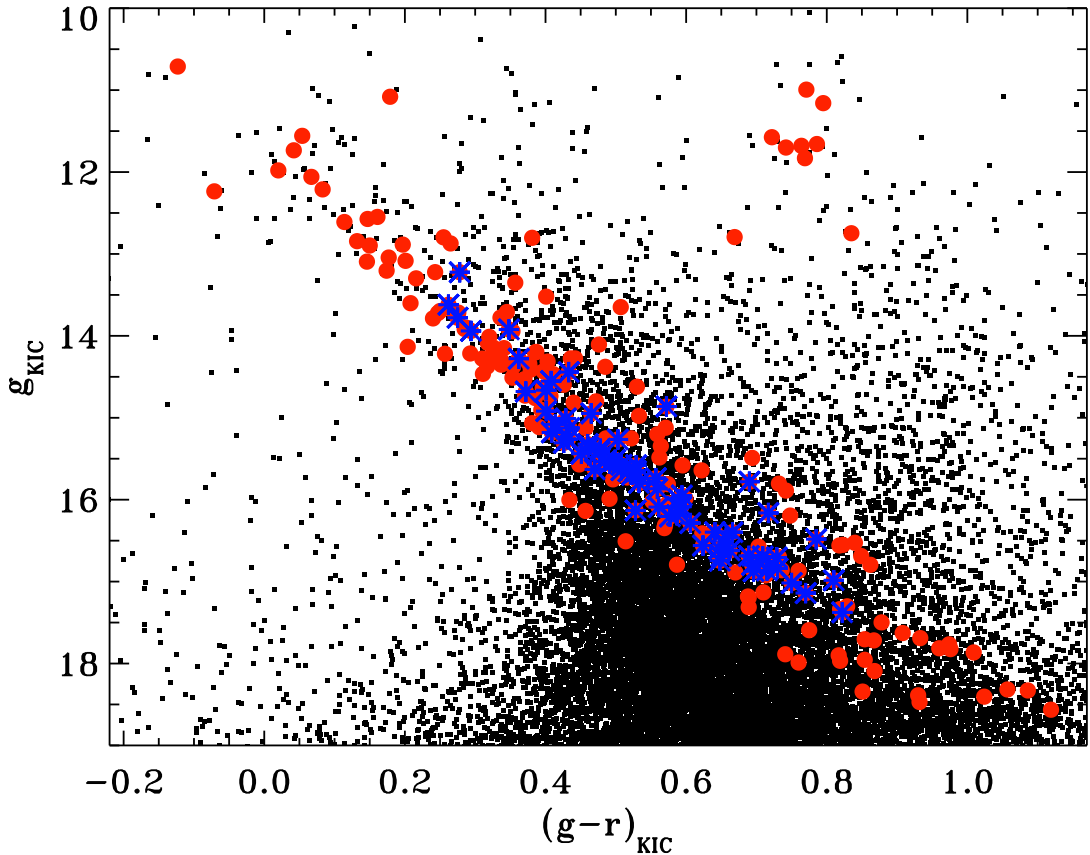}
\caption{
The color-magnitude diagram of NGC\,6811. Photometry is in
the SDSS $g$ and $r$ bands and from the Kepler Input Catalog
\citep[KIC;][http://archive.stsci.edu/kepler]{ble+11}. The
diagram shows all stars in the KIC within a 0.5-degree radius
of the cluster center. Spectroscopically single RV members and
candidate members are marked in red. The 71 members for which
we present rotation periods are marked with blue asterisks.
\label{cmd}}
\end{figure}

We are conducting a multi-epoch radial-velocity (RV) survey over a
1-degree diameter field centered on NGC\,6811 using the 6.5m MMT
telescope and the Hectochelle multi-object-spectrograph
\citep{szentgyorgyi11,mwc+07,szentgyorgyi06,ffr+05}. This work
identifies late-type members of the cluster to be observed by
{\it Kepler}. To date, $\sim$6000 spectra have been obtained of
nearly 3100 stars in the field of NGC\,6811. Of these, 363 stars
are members or candidate members\footnote{Stars with less than 4 RV
measurements are considered candidate members until additional
measurements confirm their membership.} and 228 of those have so far 
not shown significant velocity variation and are considered spectroscopically
single. These numbers underscore the high level of field star contamination
($\sim$90\% on average). Figure~\ref{cmd} shows the location of all
single members (red dots) and the 71 members for which we have
measured periods (blue asterisks) in the NGC\,6811 CMD.

\section{DATA AND ANALYSIS}
\label{data}
Details of the {\it Kepler} mission, spacecraft, and photometer have been
presented elsewhere (e.g. Borucki et al. (2010), Koch et al. (2010),
Batalha et al. (2010), Caldwell et al. (2010), Gilliland et al. (2010),
Jenkins et al. (2010a,b)). 

We analyze four quarters\footnote{The period of time between rotation of
the spacecraft around the optical axis to re-orient its solar panels and
radiator.} of {\it Kepler} data (Q1-Q4) spanning a period
of 310 days from 2009 May 12 through 2010 March 20. We use {\it Kepler}
data summed into {\it long cadence} bins ($\sim$30 minutes). Stars observed
in all 4 quarters have $\sim$13,000 flux measurements. Not all stars studied
were observed in all 4 quarters. Each stellar flux measurement is the result
of simple aperture photometry using an aperture which optimizes the
signal-to-noise ratio. The spectral response of the {\it Kepler} bandpass
(423-897nm) is similar to broadband V+R.

The stellar rotation periods presented in this paper are based on analysis
of light curves resulting from the {\it Kepler} data analysis pipeline and
corrected by the Pre-search Data Conditioning (PDC) routine. In this
process aperture photometry is performed on calibrated pixel data with
the sky signal removed. In the PDC process, attempts are made to remove
signals in the data from pointing drifts, focus changes, and thermal
variations. As described in the {\it Kepler} Data Release Notes
\citep{vcm10}, the pipeline is still under development and is primarily
intended to optimize the search for planetary transits. Accordingly, the
processing may not preserve all stellar variability with amplitudes
comparable to or smaller than the instrumental systematics on long
timescales. However, the rotation periods presented in this study are
derived from photometric variability with amplitudes (typically 1-3\%)
greater and time-scales shorter than relevant uncorrected instrumental
and data processing systematics.

\begin{figure}[!ht]
\plotone{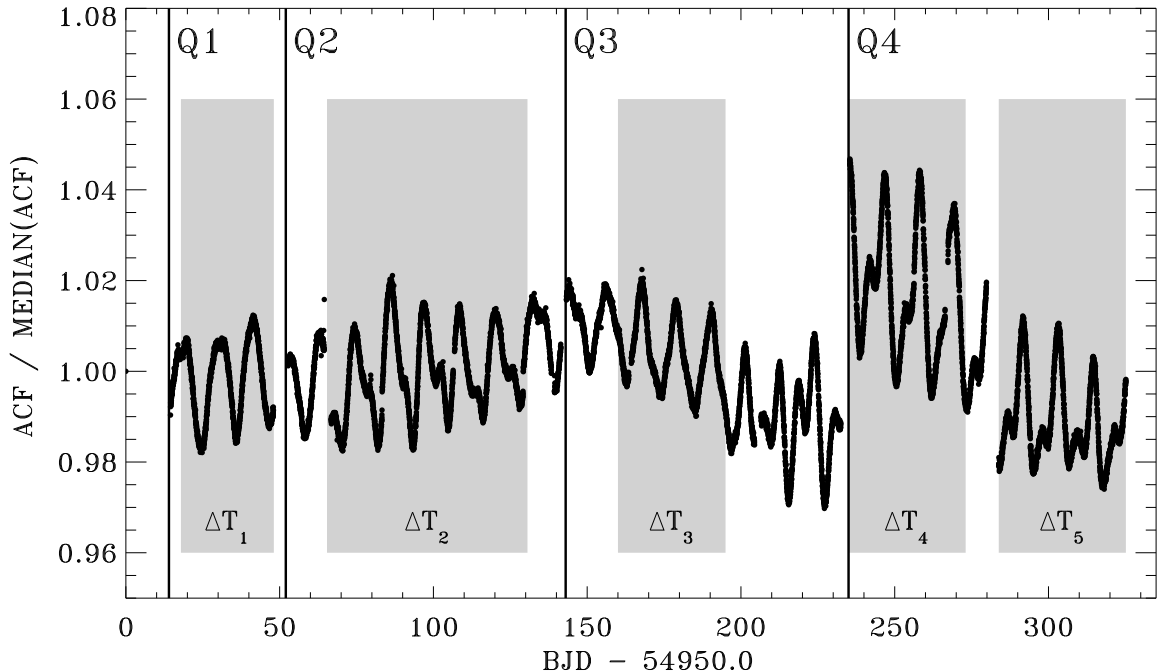}
\caption{
{\it Kepler} light curve for a single member ({\it Kepler} ID:
9531467) of NGC\,6811 showing long cadence data from Q1-Q4. The beginning
of each quarter is marked and labeled. The flux time-series for each
quarter is normalized to the median flux in the respective quarter. Grey
boxes mark the time-intervals ($\Delta T_{1}$, $\Delta T_{2}$, $\Delta
T_{3}$, $\Delta T_{4}$, and $\Delta T_{5}$) used for period determination
for this particular star.
\label{lc}}
\end{figure}

Prior to our period search, all quarters of data were normalized by the
median signal and stitched together to form a single light curve for each
star. We employed the \citet{scargle82} periodogram analysis to detect
periodic variability. We searched 20,000 frequencies corresponding to
periods between 0.05 day and 100 days. The rotation period for a given
star was determined from 2-6 time-intervals distributed over the four
quarters. These intervals were selected to avoid and minimize the effect
on the measured period from multiple spot groups and trends not removed
by the data processing. For all periods reported we have examined - by eye - 
the periodogram and raw and phased light curves. We have also checked
the periods independently using the CLEAN algorithm of \citet{rld87}.
Figure~\ref{lc} shows the normalized Aperture photometry Corrected Flux
(ACF) for a single member of NGC\,6811. In addition to the clear
modulations due to stellar spots, low frequency trends are visible for
each quarter as well as discontinuities between quarters. The five
intervals selected for measurements of the rotation period are shaded grey.

\section{RESULTS AND IMPLICATIONS}
\label{r&i}

\begin{figure}[!ht]
\plottwo{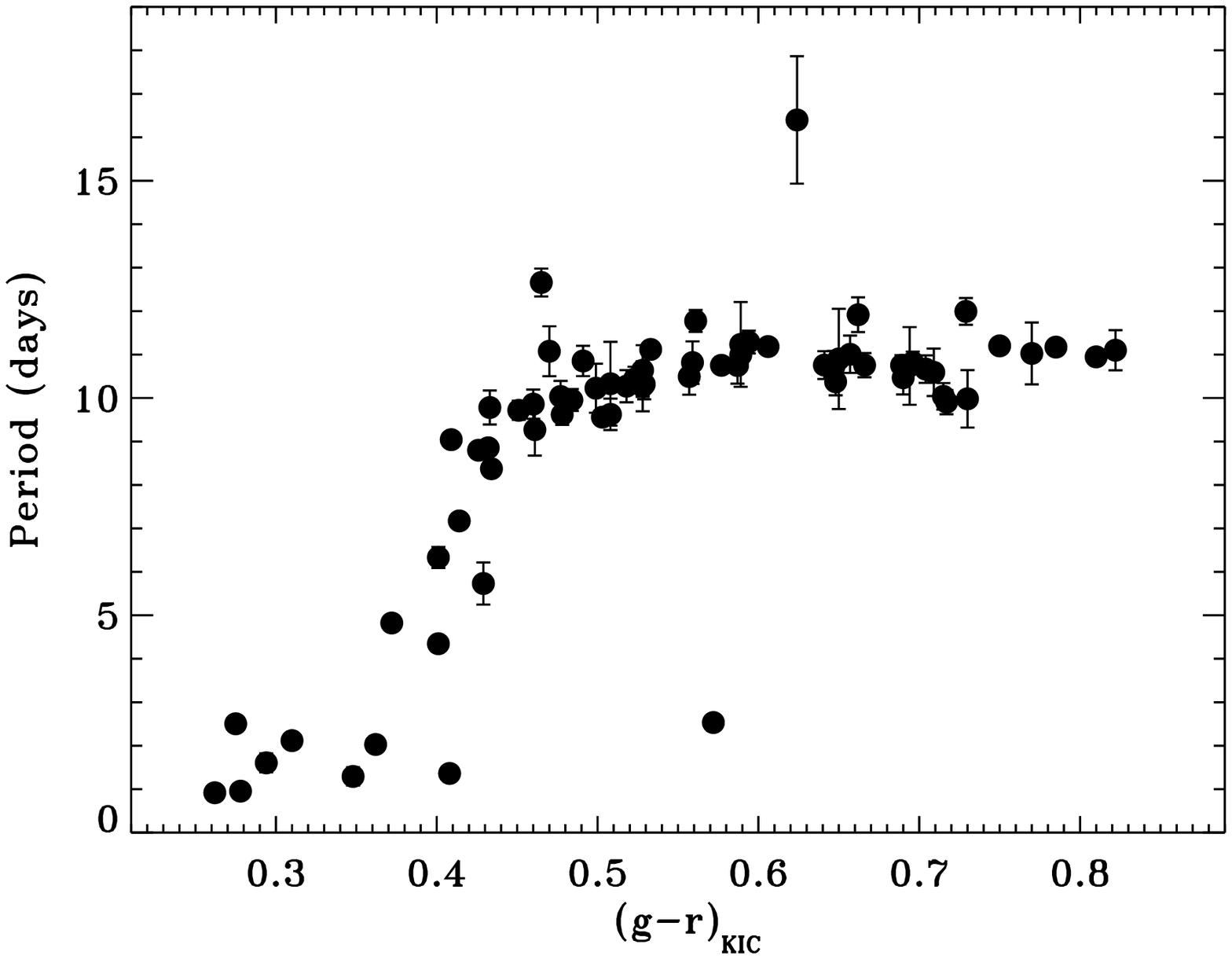}{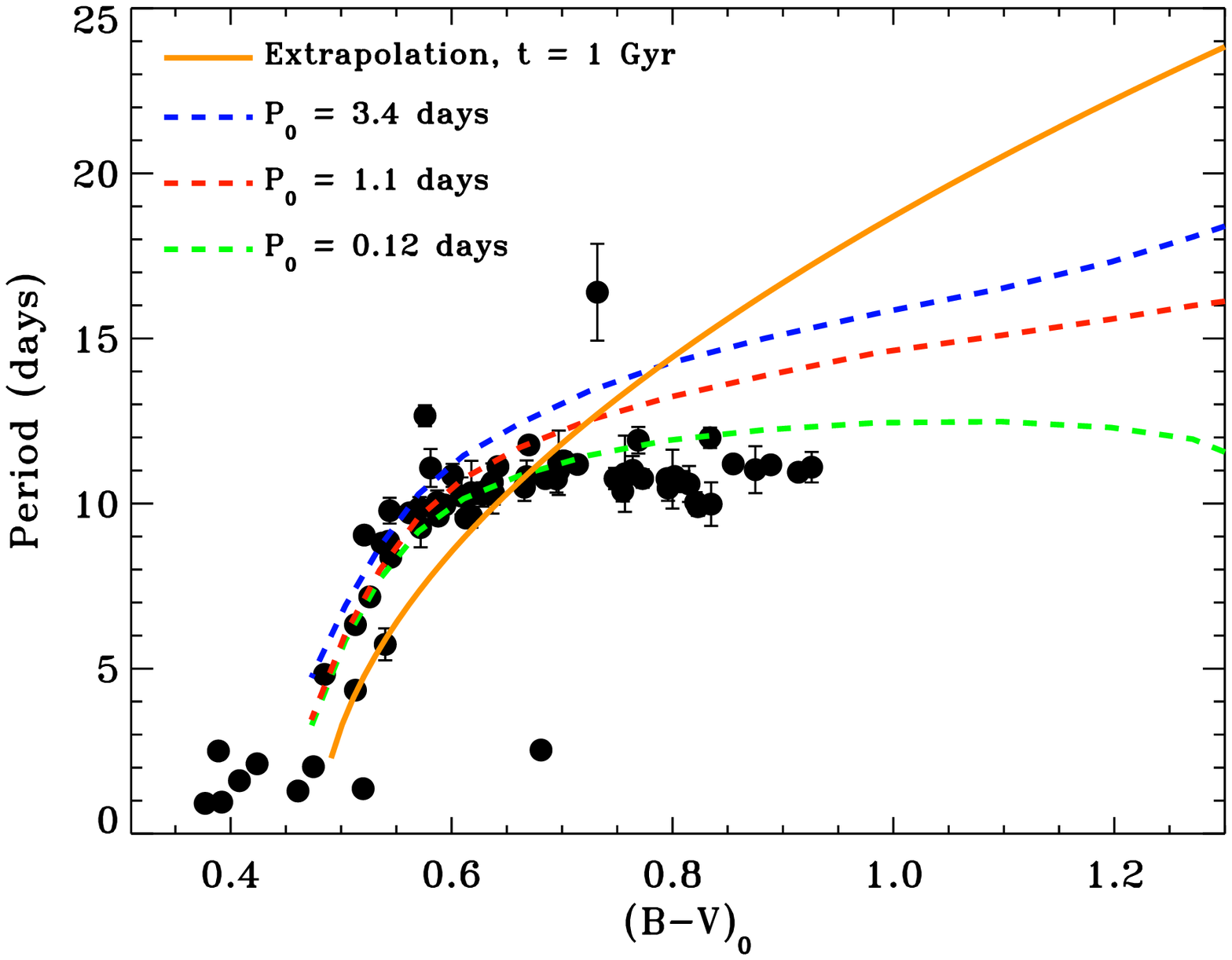}
\caption{
{\bf Left:}
The observed color-period diagram in $g$-$r$ color for 71 FGK candidate
single members of NGC\,6811. The periods define a clear $I$-type sequence.
The error-bars represent the RMS of multiple period measurements. The
sequence is the observed representation of a cross-section at the age
of NGC\,6811 ($P = P( {\rm 1\,Gyr}, M)$) of the surface $P = P(t,M)$.
{\bf Right:}
The color-period diagram in de-reddened $B$-$V$ color using $E_{(B-V)} = 0.1$.
The orange curve represents the simple extrapolation to 1\,Gyr, using the
Skumanich $P \propto \sqrt{t}$ law, of the color-period relation from younger
open clusters. The green, blue, and red curves are rotational isochrones for
$t = 1$\,Gyr calculated using the rotational evolution theory described in
\citet{barnes10}, and correspond to initial (ZAMS) periods of 0.12\,days,
1.1\,days, and 3.4\,days respectively.
\label{cpds}}
\end{figure}

The left panel of Figure~\ref{cpds} displays our color-period diagram for
NGC\,6811. Rotation periods for the 71 candidate single members identified
from the spectroscopic survey are displayed against their $g$-$r$ colors. 
We observe that, apart from a few outliers (discussed below), the
distribution of periods forms a {\it single} and {\it narrow} rotational 
sequence from mid-F through early-K spectral type. The rotation periods
are $\sim$1-2\,days for $g$-$r < 0.4$, and rise sharply at $g$-$r \sim0.4$
towards $\sim$11\,days for the redder stars. Three stars deviate significantly
from the NGC\,6811 sequence. {\it Kepler} IDs 9717386 ($P = 16.4$, $g$-$r
= 0.62$), 9595724 ($P = 12.7$, $g$-$r = 0.47$), and 9469799 ($P = 2.5$,
$g$-$r = 0.57$). All three stars are candidate single members of NGC\,6811
with 2 RV measurements and 3-4 measurements of their rotation periods over
Q1-Q4. With only 2 RV measurements their membership and/or their single
star status can change with further observations. Table~\ref{data} shows
a stub version of a table available online with Kepler IDs, astrometry,
photometry, rotation periods, and RV membership status for all 71 stars. 

\notetoeditor{This is a stub version of a table to be made available online.}

\begin{deluxetable}{lrrrrrrcc}
\setlength{\tabcolsep}{0.8mm}
\tablecaption{
Data for the 71 members of NGC\,6811 with measured rotation periods.
\label{data}}
\tablewidth{0pt}
\tablehead{
\colhead{Kepler ID} &
\colhead{RA} &
\colhead{DEC} &
\colhead{$g$} &
\colhead{$r$} &
\colhead{$P$} &
\colhead{$\sigma_{P}$} &
\colhead{$N_{P}$} &
\colhead{Class} \\
\colhead{} &
\colhead{($h~~m~~s$)} &
\colhead{($\degr~~'~~''$)} &
\colhead{} &
\colhead{} &
\colhead{(Days)} &
\colhead{(Days)} &
\colhead{} &
\colhead{}
}
\startdata
9715923  &   19 36 39.33  &   46 27 01.83  &  13.62  &  13.36  &   0.92  &   0.01  &   5  &  CSM \\
9594100  &   19 36 55.98  &   46 15 18.47  &  13.22  &  12.95  &   0.96  &   0.01  &   6  &  CSM \\
9716563  &   19 37 38.57  &   46 29 12.68  &  13.92  &  13.57  &   1.29  &   0.21  &   4  &  SM  \\
9716817  &   19 37 59.91  &   46 24 44.75  &  14.55  &  14.14  &   1.36  &   0.01  &   4  &  SM  \\
9654924  &   19 36 41.17  &   46 23 09.52  &  13.94  &  13.65  &   1.60  &   0.22  &   4  &  CSM \\
\enddata
\end{deluxetable}

To test for non-astrophysical periodic signals in the {\it Kepler} photometry
introduced by either the instrument or by the processing pipeline, we
analyzed the light curves for 100 randomly selected field dwarfs in the
NGC\,6811 field. We found no dominant frequencies and only rarely ($<5\%$)
periodic photometric variability with an amplitude similar to those
observed for members of NGC\,6811 for which we derived rotation periods. 

The sequence in the NGC\,6811 color-period diagram is equivalent to the
sequences observed for FGK dwarfs in the Hyades \citep{rtl+87,dch+11},
Praesepe \citep{dch+11}, Coma Berenices \citep{cdh+09}, and M\,37
\citep{hgp+09}, and to the slow sequences seen in numerous younger open
clusters for the same spectral range. \citet{barnes03a} labeled it the
$I$ sequence. A second sequence of ultra fast rotators with periods of
$\la$1\,day found in the youngest ($\sim$100\,Myr) open clusters [e.g.
the Pleiades \citep{va82,hbk+10} and M\,35 \citep{mms09}], is absent in
NGC\,6811 as it is in the Hyades and Coma.

The most basic implication of the NGC\,6811 color-period diagram is
that the relatively tight sequence of rotators seen in the Hyades,
Coma Berenices, Praesepe, and M\,37 continues to remain tight at later
ages, and extends to at least 1\,Gyr. Because the rotational evolution
of (single) cool stars is convergent, we expect that older clusters
will also display similarly tight sequences. The totality of such data
therefore will define the desired $P$-$t$-$M$ surface. Although based on
preliminary analysis of the spectroscopic and photometric data, the
NGC\,6811 color-period diagram constitutes a proof of concept for
the Kepler Cluster Study, and represents a significant increase
in our empirical knowledge of how cool stars spin down.

The dependence of $P$ on color (or equivalently $M$) defined by the 
NGC\,6811 $I$ sequence is a cross-section at $t \simeq 1\,Gyr$ of
the hypothesized $P$-$t$-$M$ surface. The scatter along the sequence is
consistent with the measurement uncertainty (and differential rotation)
for individual stars and confirms that the surface is intrinsically thin. 
This has important implications for its application to determine stellar
ages from measurements of colors and rotation periods \citep[gyrochronology,][]
{barnes03a,barnes07,mh08,soderblom10}. The thinner the surface the
more precisely a gyrochronology age of a star can be determined.

Knowing stellar ages is fundamental to understanding the time-evolution 
of various astronomical phenomena related to stars and their companions.
For the vast majority of stars not in clusters (unevolved late-type field
stars including most exoplanet host stars), ages determined using the
isochrone method are highly uncertain because the primary age-indicators
are nearly constant throughout their main-sequence lifetimes, and because
their distances and thus luminosities are poorly known. For such stars
rotation could serve as a superior age indicator. The tight rotational
sequence in NGC\,6811 will serve to further develop the technique of
gyrochronology and thereby enable better ages for unevolved late-type
field stars with measured periods.

Finally, we note that these results provide new and valuable constraints
for models of angular momentum evolution. The right panel of Figure~\ref{cpds}
displays the color-period diagram in de-reddened $B-V$ color [transformed
from $g$-$r$ using equations by \citet{jsr+05} and corrected by $E_{(B-V)}
= 0.1$] with the empirical color-period relation from younger clusters
extrapolated to 1\,Gyr using $P \propto \sqrt{t}$ (cross-section of orange
surface in Figure~\ref{3d}) and with rotational evolution models from
\citet{barnes10}. The green, blue, and red theoretical curves correspond
to the range of initial periods allowed on the Zero Age Main Sequence
(ZAMS). While preliminary, the comparison suggests that simple extrapolation,
using the Skumanich $\sqrt{t}$ law, of the color-period relation from
younger clusters cannot account for the shape of the NGC\,6811 sequence.
We emphasize that the current value for the cluster reddening is uncertain,
but even with agreement for late-F and early-G dwarfs, the extrapolated
and theoretical periods are longer than observed for the mid-G to early-K
dwarfs. Additional work is envisioned to understand this and other related
issues in cool star angular momentum evolution. 

Further analysis of ground-based spectroscopic and photometric data will
enable us to improve our knowledge of cluster properties such as age, distance,
and metallicity. In addition to our ongoing spectroscopic survey of the
cluster we have acquired new ground-based photometric data in UBVI filters,
and will use these to provided improved photometric information and 
reddening for the cluster in due course.

\section{CONCLUSIONS}

We present the first results from the Kepler Cluster Study for
the open cluster NGC\,6811. With rotation periods measured for 71
radial-velocity and photometric cluster members, the NGC\,6811 color-period
diagram displays a single tight rotational sequence from mid-F to early-K
spectral type. This result extends to $\sim$1\,Gyr similar sequences
observed between $\sim$550-600\,Myr in the Hyades, Praesepe, Coma
Berenices, and M\,37, and suggests that cool stars populate a thin surface in 
rotation-age-mass space. The result will enable a more detailed
understanding of the rotational evolution of cool stars and implies
that credible ages can be derived for late-type dwarfs with measured
colors and rotation periods.

\acknowledgments

Kepler was competitively selected as the tenth Discovery mission. Funding
for this mission is provided by NASA's Science Mission Directorate.
We thank the entire {\it Kepler} Mission team including engineers, managers,
and administrative staff, who have all contributed to the success of the
mission. This work has been supported by NASA grant NNX09AH18A (The Kepler
Cluster Study) to S.M., and by support to S.M. from the {\it Kepler}
mission via NASA Cooperative Agreement NCC2-1390. This paper uses data
products produced by the Optical and Infrared (OIR) Astronomy Division's
Telescope Data Center, supported by the Smithsonian Astrophysical
Observatory. S.M. express deep appreciation for time awarded on
MMT/Hectochelle and for the exceptional and friendly support of the
MMT staff. The Kepler data presented in this paper were obtained from
the Multimission Archive at the Space Telescope Science Institute (MAST).
STScI is operated by the Association of Universities for Research in
Astronomy, Inc., under NASA contract NAS5-26555. Support for MAST for
non-HST data is provided by the NASA Office of Space Science via grant
NNX09AF08G and by other grants and contracts.


{\it Facilities:} \facility{NASA {\it Kepler}}, \facility{MMT (Hectochelle)}.



\end{document}